\documentclass[twoside,11pt]{article} 
\usepackage{jmlr2e} 
\usepackage[utf8]{inputenc}
\usepackage{amsmath}
\usepackage[a4paper, margin=1in, tmargin=0.7in]{geometry}
\usepackage{verbatim}
\usepackage{natbib}
\AtBeginDocument{%
 \abovedisplayskip=7pt plus 2pt minus 2pt
 \abovedisplayshortskip=0pt plus 1pt
 \belowdisplayskip=7pt plus 2pt minus 2pt
 \belowdisplayshortskip=4pt plus 1pt minus 1pt
}

\title{How to measure the topological quality of protein grammars?}
\author{\name Witold Dyrka \email witold.dyrka@pwr.edu.pl\\ 
\addr Politechnika Wrocławska, Wydział Podstawowych Problemów Techniki, Katedra Inż. Biomedycznej
\AND 
\name Fran\c{c}ois Coste\\ 
\addr Irisa / Inria Rennes - Bretagne Atlantique, France
\AND
\name Olgierd Unold,
\name Łukasz Culer,
\name Agnieszka Kaczmarek\\
\addr Politechnika Wrocławska, Wydział Elektroniki, Katedra Informatyki Technicznej, Poland
} 

\date{}

\begin{document}
\pagenumbering{gobble}

\maketitle

\let\thefootnote\relax\footnote{The research project has been partially funded by National Science Center (grant no. 2015/17/D/ST6/04054). Some calculations have been carried out using resources provided by Wroclaw Centre for Networking and Supercomputing (http://wcss.pl), grant No. 98.}

\vspace{2mm}

\paragraph{Motivation.}
Context-free (CF) and context-sensitive (CS) formal grammars are often regarded as more appropriate to model proteins than regular level models such as finite state automata and Hidden Markov Models (HMM). In theory, the claim is well founded in the fact that many biologically relevant interactions between residues of protein sequences have a character of nested or crossed dependencies. In practice, there is hardly any evidence that grammars of higher expressiveness have an edge over old good HMMs in typical applications including recognition and classification of protein sequences. This is in contrast to RNA modeling, where CFG power some of the most successful tools. There have been proposed several explanations of this phenomenon. On the biology side, one difficulty is that interactions in proteins are often less specific and more "collective" in comparison to RNA. On the modeling side, a difficulty is the larger alphabet which combined with high complexity of CF and CS grammars imposes considerable trade-offs consisting on information reduction or learning sub-optimal solutions. Indeed, some studies hinted that CF level of expressiveness brought an added value in protein modeling when CF and regular grammars where implemented in the same framework \citep{Dyrka07, Dyrka13}. However, there have been no systematic study of explanatory power provided by various grammatical models. The first step to this goal is define objective criteria of such evaluation. Intuitively, a decent explanatory grammar should generate topology, or the parse tree, consistent with topology of the protein, or its secondary and/or tertiary structure. In this piece of research we build on this intuition and propose a set of measures to compare topology of the parse tree of a grammar with topology of the protein structure. 


\paragraph{Measures.}
First, we define two measures, inspired by the silhouette plot, that relate path lengths in the parse tree between residues in contact to path lengths between residues \textit{not} in contact. Let 
\begin{equation}
    d_{C} = \frac{1}{|PC^L|}\sum_{(i,j) \in PC^L} { d(i,j) }
    \mathrm{\quad and \quad}
    d_{NC} = \frac{1}{|PNC^L|}\sum_{(i,j) \in PNC^L} { d(i,j) },
\end{equation}
where $PC^L$ ($PNC^L$) is the set of pairs of residues $i,j, |i-j|\geq L$ in contact (\textit{not} in contact), and $d(i,j)$ is the length of the shortest path between $i$ and $j$ in the parse tree. Only pairs of residues s.t. $|i-j|\geq L$ are included in $PC^L$ and $PNC^L$. Then, we define two closely related measures of the overall fitness of the parse tree w.r.t. to a given contact map:
\begin{equation}
    S_1 = \frac{d_{NC}-d_{C}}{max\{d_{C},d_{NC}\}},
    \mathrm{\quad and \quad}
    R_1 = \frac{d_{NC}}{d_{C}}.
\end{equation}
Respectively, they are a normalized difference and a ratio of average path lengths between residues in contact and between residues \textit{not} in contact. $S_1$ gives scores from $-1$ (\textit{bad}) to $1$ (\textit{good}), while $R_1$ ranges from $0$ (\textit{bad}) to $\infty$ (\textit{good}). Both measures indicate how much paths to residues in contact are shorter than paths to residues \textit{not} in contact. 

In addition, local variants of the measures defined by restricting $PC^L$ ($PNC^L$) to pairs involving a residue $i$ can be used to assess placement of the residue $i$ in the parse tree.

Finally, we propose to measure overlap between a subset of residues close in a protein structure and a subset of residues close in a parse tree:
\begin{equation}
    D_1 = 2 \times \frac{|PP^L(t) \cap PC^L|}{|PP^L(t)| + |PC^L|},
\end{equation}
where $PP^L(t)$ is the subset of all pairs of residues $i,j, |i-j|\geq L$, s.t. the length of the shortest path between them in the parse tree is less than $t$. Note that $D_1$ is identical to the Dice coefficient. 
In addition, for weighted grammars, we suggest to use variants of the measures where the length of the shortest path in the parse tree $d(i,j)$ is substituted by the logarithm of the product of edge weights between $i$ and $j$.

\paragraph{Early Results.}We applied the proposed measures to evaluate topology of grammars generated for the HET-s prion forming domain from fungi. The main part of the HET-s fold is a loop-like structure with a nested pattern of dependencies between hydrophobic residues, which makes the fold a suitable target for modeling using a CFG. The positive sample was a non-redundant subset (similarity$<$70\%) of 21 amino acid-long HET-s-related motifs r1 and r2 identified in a recent study \citep{Daskalov15} (156 cases). The negative sample was a non-redundant subset of all 21 amino acid-long fragments from the negative sample used by \citet{Dyrka09} (5760 cases). A contact map of the HET-s structure was obtained from the 2rnm PDB entry \citep{Wasmer08} using the most typical $C_\beta$-$C_\beta$ contact cutoff of 8\AA\  and sequence separation $L$ of 5 residues.

As a starting point, two reference parse trees were hand-crafted based on the HET-s contact map: one consistent with the Chomsky Normal Form (\textit{hc/cnf}) and the other in the free form (\textit{hc/ff}). According to our measures, the trees obtained the following scores: $\textit{hc/ff: }$ ${S_1=0.54}$, ${R_1=2.16}$, ${D_1(t=5)=0.77}$ and $\textit{hc/cnf: }$ ${S_1=0.46}$, ${R_1=1.84}$, ${D_1(t=8)=0.36}$. This confirms that parse tree topologies generated by CFG may convey information regarding the protein structure. In the next step, CFG were trained in the 4-fold cross-validation scheme using two existing frameworks: non-probabilistic Grammar-based Classifier System (GCS) \citep{Unold05, Unold07} which learned from positive and negative samples and probabilistic Protein Grammar Evolution (PGE) \citep{Dyrka09,Dyrka13} which learned from positive samples and assumed a starting grammar topology \textit{NestedNT} originally proposed for binding sites \citep{Dyrka09}. In addition, a combined approach was used s.t. PGE learned probabilities for a set of rules obtained by GCS, slightly modified to fulfill PGE requirements. Probabilistic grammars performed well in the classification test with average \textit{F1} up to over 0.4, a good result providing the ratio of positive and negative sample sizes reaching 1:37. A non-probabilistic grammar obtained \textit{F1} of 0.07. However, these performances did not translate to quality of topologies of most likely parse trees generated by the grammars: for most of them average values of the three measures calculated for the positive test samples were close to random levels (${S_1\sim0.0}, {R_1\sim1}, {D_1<0.05}$). The most notable exception was a charge property-based (see \citep{Dyrka09}) probabilistic CFG generated by PGE for one of the folds, which reached ${S_1=0.33}, {R_1=1.52}, {D_1=0.25}$. The best probabilistic CFG generated by PGE based on GCS-derived topology achieved ${S_1=0.17}, {R_1=1.26}, {D_1=0.09}$.

The preliminary tests seem to support usefulness of the proposed measures for assessing consistency of parse trees with topology of proteins. In addition, our early results suggest that learning grammars that represents protein topologies may require dedicated approaches. More tests are needed to draw solid conclusions, while the line of research appears very promising.

\bibliographystyle{plainnat}
\bibliography{references}

\end{document}